\documentclass{llncs}
\usepackage{amsmath}
\usepackage{hyperref}

\title{ Project Paper: Embedding generic monadic transformer into Scala }
\subtitle{ Can we merge monadic programming into mainstream? }
\author{ Ruslan Shevchenko 
 \orcidID{ 0000-0002-1554-2019 }
}
\institute{
  \email{ ruslan@shevchenko.kiev.ua }
}

\usepackage{listings}
\usepackage{stmaryrd}
\usepackage{bigfoot}

\begin{document}

\maketitle

\begin{abstract}
  Dotty-cps-async is an open-source package that consists of scala macro, which implements generic async/await via monadic cps transform, and library, which provides monadic substitutions for higher-order functions from the standard library. It allows developers to use direct control flow constructions of the base language instead of monadic DSL for various applications. Behind well-known async/await operations, the package provides options for transforming higher-order function applications, generating call-chain proxies, and automatic coloring.
\end{abstract}

\section{ Introduction }

  It's hard to have evidence-based data about industry adoption.  From the subjective observations,  one of the barriers to industrial adoption of the Scala language is an unnecessary high learning curve. 
 The tradition of using embedded DSL ({\it Hudak} \cite{10.1145/242224.242477}) instead of the base language leads to a situation when the ‘cognitive load’  of relatively simple development tasks, such as querying an extra resource, is higher than in mainstream languages. A programmer cannot use control flow constructions of base language but should learn a specific DSL and use a suboptimal embedding of this DSL, usually within monadic for comprehensions. Therefore, developers proficient in Java or TypeScript cannot be immediately proficient in Scala without additional training.

 Can we provide a development environment that gives the programmer an experience comparable to the state of the art mainstream back-end programming, such as Kotlin structured concurrency, Swift async functions or \verb|F#| computation expressions ? 
 Dotty-cps-async intends to be an element of the possible answer.  It provides the way to embed monadic expressions into base scala language using well-known async/await constructs, existing for nearly all mainstream programming languages.  Although the main idea is not new, dotty-cps-async provides behind well-known interfaces a set of novel features,  such as support of the generic monads, transformation of higher-order function applications, generation of call-chain proxies, and automatic coloring.   

  The package is open-source and can be downloaded from the GitHub repository {\small \verb|https://github.com/rssh/dotty-cps-async|}.
  
    The paper is organized as follows: In section  {\ref{EmbeddingGeneric}} we briefly describe the async/await interface and the process of monadification; in \ref{MonadsParametrization} add monads parametrization and provide some examples of applying async/await transformation with non-standard monads; in \ref{HO} describe a support of using await inside the arguments of higher-order functions, then in \ref{AutomaticColoring} intriduce an optional automatic colouring facility and show an example of non-blocking file copying on \ref{AACopyFile} on page \pageref{AACopyFile}  elaborated with coloring on page \pageref{AACopyFileAC}.  
    Short overview on related work in this area is in section \ref{RelatedWork}.

\section{ Embedding generic monadic cps transform into Scala  }  \label{EmbeddingGeneric}

Dotty-cps-async implements an interface similar to scala-async by ({\it Haller}  \cite{hallerScalaAsync}) based on optimized monadic CPS transform.  It is implemented as Scala macros and provides a simple generic interface with a well-known async/await signature, slightly changed to support monad parametrization.

In simplified form this is a combination of two generic pseudofunctions:

\begin{lstlisting}
     def async[F[_],T](f:  T): F[T] 
 \end{lstlisting}

\begin{lstlisting}
     def await[F[_],T]( x: F[T]):  T 
 \end{lstlisting}

where we can use \lstinline|await| inside the \lstinline|async| block .

Complete definitions are more complex, although  most of this complexity is hidden from the application programmer.
    Let's look at it in detail and briefly describe the basics of the Scala 3 language features used in those definitions:
 
 Full definition of \lstinline|async| :
    
\begin{lstlisting}
 transparent inline 
     def async[F[_]](using monad:CpsMonad[F])
                                      = InferAsyncArg(monad)
     
 class InferAsyncArg[F[_]](am:CpsMonad[F]) {
    transparent inline def apply[T](f: am.Context ?=> T): F[T] 
 }
 \end{lstlisting}

\begin{itemize}
 \item \lstinline|F[_]| is a type parameter of async. The compiler will deduce type parameters automatically from context if it is possible.  Underscore inside square brackets in \lstinline|F| means  F is a higher-kinder type with one type parameter.
 \item \lstinline|transparent inline| means a macro, which should be expanded during typing phase of the compiler.The compiler passes the typed tree to the macro and refines the output type of the result of the macro application.
 \item \lstinline|using| clause at the beginning of the argument list, means that the compiler will substitute the appropriative parameters by the given instance of the parameter type,  if such instance is defined in the current scope.  
 Given instance can be introduced in scope via \lstinline|given| clause.  For example, the following definition  will introduce \lstinline|myMonad|. as given instance of \lstinline|CpsMonad[Future]|.
\begin{lstlisting}
given CpsMonad[Future]  = myMonad
\end{lstlisting}
  
  Exists predefined function \lstinline|summon[A]| which will give us a given instance of A if it's defined.
  
 \item Our \lstinline{async} returns an object of the auxiliary class
    \lstinline{InferAsyncArg} that defines \lstinline{apply} method.  In Scala,
    any object with the \lstinline{apply} method can be applied as if it were
    a function. The auxiliary class is a work-around for the lack of
    multiple type parameters in function definitions. The ideal
    definition, without auxiliaries, would have looked as:
\begin{lstlisting}
 transparent inline 
     def async[F[_]](using am:CpsMonad[F])[T](am.Context ?=>T)
\end{lstlisting}
The optimizer should erase the constructor of the intermediate class.  
  \item  \lstinline| f: (am.Context ?=>> T)] | is a context function. Here  \lstinline|f| is an expression of type \lstinline|T| in context  \lstinline|am.Context|.  Inside  context function we can use \lstinline|summon[am.Context]| to access a context parameter.
  Context provides an API that can be used only in \lstinline|async| block.
  \item \lstinline| am.Context | is a path-depended type.  Type \lstinline|Context| should be defined in the scope of \lstinline|am|.
\end{itemize} 

When the developer uses async pseudo function in the application code, the compiler will pass to the macro transformation typed tree with expanded implicit parameters and context arguments.

An example of original code:
\begin{lstlisting}
async[Future] {
   val ec = summon[FutureContext].executionContext
    val x = doSomething()
    ....
}
\end{lstlisting}

Expansion passed to the macro:
\begin{lstlisting}
async[Future](using FutureAsyncMonad()).apply{ (fc:FutureContext) ?=>
   val ec = fc.executionContext
   val x = doSomething()
   .....
}
\end{lstlisting}

Here we assume that \lstinline|FutureAsyncMonad|. is a given monadic interface for \lstinline|Future| where type \lstinline|Context| is defined as alias to \lstinline|FutureContext|.

Monadic operations are defined in the given \lstinline|CpsMonad[F]| parameter which should implement the following typeclass:

\begin{figure}
\begin{lstlisting}
 trait CpsMonad[F[_]] {
   type Context <: CpsMonadContext[F]
   def pure[A](v:A): F[A]
   def map[A,B](fa:F[A])(f: A=>B): F[B]
   def flatMap[A,B](fa:F[A])(f: A=>F[B]): F[B]
 }
\end{lstlisting}
\end{figure}

Optionally extended by error generation and handling operations:

\begin{figure}
\begin{lstlisting}
trait CpsTryMonad[F[_]] extends CpsMonad[F] {
  def error[A](e: Throwable): F[A]
  def flatMapTry[A,B](fa:F[A])(f: Try[A] => F[B]):F[B]
}
\end{lstlisting}
\end{figure}

The \lstinline|Cps| prefix here refers to the relation with continuation passing style for which transformation from direct to monadic style is closely related. This prefix is more related to a library rather than an interface:  \lstinline|dotty-cps-async| should coexist with generic functional programming frameworks, like \lstinline|cats| or \lstinline|scalaz| where other forms of \lstinline|Monad| are defined, so we need to have another name, to prevent name resolution conflict in application code. 

Now let us look at the full definition of \lstinline|await|:

\begin{lstlisting}
  def await[G[_],T,F[_])(x:G[T])
                      (using CpsMonadConversion[G,F], 
                                 CpsMonadContext[F]):T
\end{lstlisting}

Here \lstinline|G|. is a type which we awaiting, \lstinline|F| -- monad which is used in enclosing \lstinline|async|.
Note that $F$ and $G$ can be different; if the given instance of conversion morphism from $G[\_]$ to $F[\_]$ is defined in the current scope, then \lstinline|await[F]| can be used inside \lstinline|async[G]|.

 This morphism is represented by the \lstinline|CpsMonadConversion| interface:
 \begin{lstlisting}
trait CpsMonadConversion[G[_],F[_]] {
  def apply[T](gt:G[T]): F[T]
}
\end{lstlisting}

\lstinline|CpsMonadContext[F]| is an upper bound of \lstinline|CpsMonad[F].Context| with one operation defined:

 \begin{lstlisting}
trait CpsMonadContext[F[_]] {
    def adoptAwait[T](v:F[T]):F[T] 
}
\end{lstlisting}
  The work of \lstinline|adoptAwait| is to pass information from the current monad context into the awaiting monad.  For example, this can be a cancellation event in the implementation of a structured concurrency framework.

  Underlying source transformation is an optimized version of monadification (Ervig, Martin and others \cite{10.1016/j.scico.2004.03.004} ), similar to translating terms into continuations monad (Syme \cite{10.1145/174675.178053}). This translation is limited to the code block inside an async argument.  

We will use notations $F.op$ as shortcut for appropriative operation over monad typeclass for $F$. 
$C_F \llbracket code  \rrbracket $ is a translation of code $code$ in the context of $F$.

Let us recap the basic monadification transformations adopted to scala control-flow construction:

$$
\begin{array}{ l l }\\
   \text{trivial}  &  \frac{ C_F \llbracket t \llbracket  \textrm{ where  $t$ ia a constant or an identifier }} { F.pure(t) } \\
\\
   \text{sequential} & \frac{ C_F \llbracket \{a;b\} \rrbracket }
                                        { F.flatMap(C_F \llbracket a \rrbracket )(\_ \Rightarrow C_F \llbracket b \rrbracket ) } \\
\\
   \text{val definition} & \frac{ C_F \llbracket {val\,a = b;\, c} \rrbracket}
                                            { F.flatMap(C_F \llbracket a \rrbracket )(a' \Rightarrow C_F \llbracket b_{[a \leftarrow a']} \rrbracket) } \\
\\
   \text{condition} & \frac{ C_F \llbracket if\,a\,then\,b\,else\,c \rrbracket }
                       { F.flatMap(C_F \llbracket a \rrbracket)(a' \Rightarrow if\,(a')\,then\,C_F \llbracket b \rrbracket \,else\,C_F \llbracket c \rrbracket ) } \\
\\
   \text{match} & \frac{ C_F \llbracket a\,match\{\,case\,\, r_1 \Rightarrow v_1 \dots r_n \Rightarrow v_n\} \rrbracket }
     {F.flatMap(C_F \llbracket a \rrbracket)\{a' \Rightarrow 
         match\{ case\, r_1 \Rightarrow C_F \llbracket v_1 \rrbracket \dots r_n \Rightarrow C_F \llbracket v_n \rrbracket \} \} } \\
\\
 \text{while} & \frac{C_F \llbracket while(a)\{ b \} \rrbracket }
                              {whileHelper(C_F \llbracket a \rrbracket ,C_F \llbracket b \rrbracket )} \\
 \\
\end{array}
$$
 where $whileHelper$ is a helper function, defined as 
\begin{lstlisting}
def whileHelper(cond: F[Boolean], body: F[Unit]):F[Unit] =
   F.flatMap(cond){ c =>
      if (c) {
         F.flatMap(body){ _ => whileHelper(cond, body) }
      } else {
         F.pure(())
      }
   }
\end{lstlisting}

$$
\begin{array}{l l} \\
 \text{try/catch}  & \frac{C_F \llbracket try\{a\}\{ catch\, e \Rightarrow b \}\{ finally\, c \} \rrbracket }
                     {
                      \mbox{\tiny{\ensuremath{
                       \begin{array}{l}
                         F.flatMap( \\
                         \,\, F.flatMapTry(C_F \llbracket a \rrbracket)\{ \\
                         \,\,\,\,\,\,case\,Success(v) => F.pure(v) \\
                         \,\,\,\,\,\,case\,Failure(e) => C_F \llbracket b \rrbracket \\
                         \,\,\} \\
                         )\{ x \Rightarrow F.map(C_F \llbracket c \rrbracket ,x) \} \\
                       \end{array}
                      } } }
                     }  \\
\\
\\
 \text{throw} & \frac{ C_F \llbracket throw\,ex \rrbracket } 
                              { F.error(ex) } \\
\\
 \text{lambda} &  \frac{C_F \llbracket a \Rightarrow b \rrbracket }
                                  {a \Rightarrow C_F \llbracket b \rrbracket }.  \\
\\
 \text{application} &
            \frac
            { C_F \llbracket f(a) \rrbracket  \textrm{ where  $a$  is non-functional type} }
            { F.flatMap(C_F \llbracket a \rrbracket )(x => f(x)) } \\
\\
                   &
            \frac       
            { C_F \llbracket f(a) \rrbracket  \textrm{where $f$ is a lamba-function. } } 
            {F.flatMap(C_F \llbracket a \rrbracket )(x => C_F \llbracket f \rrbracket (x)) } \\
\\
                  &
            \frac      
            { C_F \llbracket  f(a) \rrbracket  \textrm{ where }   \exists f'  \textrm{ is an external-provided shifted variant of $f$ } } 
            { F.flatMap(C_F \llbracket  a \rrbracket  )(x => f'(x))}  \\
\\

\text{await} &
           \frac { C_F \llbracket await_G(a)(m,c) \rrbracket   \textrm{ where }  F = G  }{ c.adoptAwait(a)  } \\
\\
           &
           \frac{ C_F \llbracket  await_G(a)(m,c) \rrbracket   \textrm{ where }  F \ne G  }{ c.adoptAwait(CpsMonadConversion[F,G](a))  }  \\
\\
                    
\end{array}
$$

The mechanism for definition and substitution of shifted functions is described in \ref{HO}.

Implementation differs from the basic transformation, by few optimizations, which are direct applications of unit monad law: 
\begin{itemize}
\item few sequential blocks with trivial CPS transformations are merged into one:
$$
    {F.flatMap(F.pure(a))(x \Rightarrow F.pure(b(x))} \over { F.pure(b(a)) }
$$
 \item translation of control flow operators are specializated for the cases when transformations of some subterms are trivial. In the best case, control-flow construction is lifted inside monad barriers.   For example, the transformation rules for \lstinline|if-then-else| taking into account optimizations will look like:
$$
\begin{array}{ l l }\\
   \,\,\,\,\,\, & \frac
                  { C_F \llbracket  if\,a\,then\,b\,else\,c \rrbracket  , 
                        C_F \llbracket  a \rrbracket  \ne F.pure(a) \land C_F \llbracket  b \rrbracket  \ne F.pure(b) \land 
                        C_F \llbracket  c \rrbracket  \ne F.pure(c) } 
                  { F.flatMap(C_F \llbracket  a \rrbracket  )(v => if (v) then\, C_F \llbracket  b\rrbracket \, else\, C_F \llbracket c \rrbracket ) } \\
   \\
   \,\,\,\,\,\, & \frac
                   { C_F \llbracket  if\,a\,then\,b\,else\,c \rrbracket , 
                     C_F \llbracket  a \rrbracket  = F.pure(a) \land 
                        (C_F \llbracket b \rrbracket  \ne F.pure(b) \lor C_F \llbracket  c \rrbracket  \ne F.pure(c)) } 
                  { if\,a\,then\,C_F \llbracket  b \rrbracket \,else\,C_F \llbracket  c \rrbracket  } \\
   \\
   \,\,\,\,\,\, & \frac 
                  { C_F \llbracket if\,a\,then\,b\,else\,c \rrbracket , 
                      C_F \llbracket  a \rrbracket  = F.pure(a) 
                      \land C_F \llbracket  b \rrbracket  = F.pure(b) 
                      \land C_F \llbracket c \rrbracket  = F.pure(c) } 
                  { F.pure(if\, a\, then\, b\, else\, c) } \\
\\
\end{array}
$$
  Such specializations are defined for each control-flow construction. 

 \end{itemize}

 In the resulting code, the number of monadic binds is usually the same as a number of awaits in the program, which made performance characteristics of code, written in a direct style and then transformed to the monadic style,  the same, as the code manually written by hand in monadic style.

\subsection{ Monads parametrization }. 
\label{MonadsParametrization}

  Async expressions are parameterized by monads, which allows the CPS macro to support behind the standard case of asynchronous processing other more exotic applications, such as processing effects( {\it Syme } \cite{10.1145/601775.601776}), ({\it Brachth{\"{a}}user} \cite{DBLP:journals/jfp/BrachthauserSO20}),  logical search  ({\it Kiselyov and Shan } \cite{10.1145/1090189.1086390}), or probabilistic programming ({\it Adam and Ghahramani and others}\cite{10.1145/2887747.2804317}). 
Potentially,  list of problem domains from \verb|F#| computation expression Zoo ({\it Tomas and Syme} \cite{computation-zoo-padl14}) is directly applicable to dotty-cps-async.

 In practice, most used monads constructed over \verb|Future|, effect wrappers, like IO and constructions over effect classes extended by additional custom logic.

Let's look at the following example:

\label{AACopyFile}
\begin{lstlisting}
 val prg = async[[X] =>> Resource[IO,X]] {
   val input = await(open(Paths.get(inputName),READ))
   val output = await(open(outputName,WRITE, CREATE, TRUNCATE_EXISTING))
   var nBytes = 0
   while
      val buffer = await(read(input, BUF_SIZE))
      val cBytes = buffer.position()
      await(write(output, buffer))
      nBytes += cBytes
      cBytes == BUF_SIZE
   do ()
   nBytes
 }
\end{lstlisting}

 Here \lstinline|[X] =>> Resource[IO,X]| is a type-lambda, which represent the computational effect monad IO,  extended by the acquisition and release of resources of type of the argument \lstinline|X|.

 Inside \lstinline|async[[X] =>> Resource[IO,X]]|,  \lstinline|input| and \lstinline|output|. resources will be automatically closed at the end of the appropriative scope.

 Without async/await, one would have to write the following:

\begin{lstlisting}[basicstyle=\small]
 (
  for{
   input <- open(Paths.get(inputName),READ)
   output <- open(outputName,WRITE, CREATE, TRUNCATE_EXISTING)
  } yield (input, output)
 ).evalTap{ case (input, output) =>
    var nBytes = 0
    def step(): IO[Unit] = {
      read(input, BUF_SIZE).flatMap{ buffer =>
        val cBytes = buffer.position()
        write(output, buffer).flatMap{ _ =>
           nBytes += cBytes
           if (cBytes == BUFF_SIZE) 
              step()
           else
              IO.pure(())
        }
      }
    }
    step().map{ _ => nBytes }
 }
\end{lstlisting}

\label{CombSearch}

 The next example illustrate a monadic representation of combinatorial search. Monad 
   \lstinline|[X] =>> ReadChannel[Future,X]| represent a CSP[Communicating Sequential Processes]-like channel {\it Hoare} \cite{hoare1985communicating}, 
  where monadic combinators apply the functions over the stream of possible states.

We want to solve the classical N-Queens puzzle: placing N queens on a chessboard so that no two figures threaten each other.  

Let us represent the chessboard state as a vector of queens second coordinates $queens$ with additional helper method \lstinline|isUnderAttack| with obvious semantics.  The $i$-th queen is situated at $(i,queens[i])$ location.

\begin{lstlisting}
  type State = Vector[Int]
     
  extension(queens:State) {

    def isUnderAttack(i:Int, j:Int): Boolean = 
      queens.zipWithIndex.exists{ (qj,qi) => 
        qi == i || qj == j || i-j == qi-qj || i+j == qi+qj
      }
      
    def asPairs: Vector[(Int,Int)] =
     queens.zipWithIndex.map(_.swap)

  }
\end{lstlisting}

 Function \lstinline|putQueen|. generate from one starting state a channel of possible next states.
 \lstinline|async[Future]| in \lstinline|putQueen| spawns a concurrent process for enumerating the next possible steps in N-Queens solution.  

\begin{lstlisting}
 def  putQueen(state:State): ReadChannel[Future,State] =
   val ch = makeChannel[State]()
   async[Future] {
     val i = state.length
     if i < N then
       for{ j <- 0 until N  if !state.isUnderAttack(i,j) }
         ch.write(state appended  j )
     ch.close()
   }
   ch
\end{lstlisting}

And we can recursive explore all possible steps with help of the \verb|solution| function, which returns the stream of finish states:

\begin{lstlisting}
  def solutions(state: State): ReadChannel[Future,State] =
   async[[X] =>> ReadChannel[Future,X]] {
     if(state.length < N) then
       val nextState = await(putQueen(state))
       await(solutions(nextState))
     else
       state   
   }
\end{lstlisting}

(runnable version is available at \url{https://github.com/rssh/scala-gopher/blob/master/shared/src/test/scala/gopher/monads/Queens.scala})

The computation is directed by reading from the stream of solutions.  
In \lstinline|putQueen|, the computations inside the loop will suspend after each write to the output channel, until all descendent states are explored.
 The suspension point \lstinline|await| is hidden inside \lstinline|ch.write| inside for loop.

\lstinline|ch.write| is defined in \lstinline|ReadChannel[F,A]| as
\begin{lstlisting}
 transparent inline def write(inline a:A): Unit = 
      await(awrite(a))(using CpsMonad[F])
\end{lstlisting}
 transparent inline macros in scala are expanded in code at the same compiler phase before enclosing macro, so async code transformer process this expression in for loop instead \lstinline|ch.write|.
 
In such way
 \lstinline|solutions(State.empty).take(2)|  will return the first two solutions without performing a breadth-first search. 

\subsection{ Translation of higher-order functions }

\label{HO}

Supporting cps-transformation of higher-order functions is important for a functional language because it allows \lstinline|await| expression inside loops and arguments of common collection operators.
  As an example, in the previous section await inside \lstinline|for| loop was used for asynchronous channel write. Using await inside higher-order function enables idiomatic functional style, such as
\begin{lstlisting}
  val v = cache.getOrElse(url, await fetch(url)) 
\end{lstlisting}
  
   Local cps transform changes the type of a lambda function.  If the runtime platform supports continuations, we can  keep the shape of the arguments in the application unchanged by defining 'monad-escape' function transformers, which can restore the view of $cps(f):A \Rightarrow F[B]$ back to $A \rightarrow B$.  

However, for the platform without continuation support, higher-order functions from other module is a barrier for local async transformations.   For the time of the writing of this article, none of the available Scala runtimes (i.e. JVM, Js, or Native) have continuations support.  For JVM exists a plan to implement continuations support via Project Loom \cite{ProjectLoom},  but it is not available for production use yet.  JavaScript runtime is significantly smaller than JVM, and runtime semantics is precisely defined asynchronous.  

  For those runtimes and for cases when semantic of monad does not allow us to build such escape function, dotty-cps-async implements limited support of higher-order functions.  Macro performs a set of transformations, which allows developers to describe the substitution for the origin higher-order function in their code.

Let us have a first-order function: $f: A \Rightarrow B$ which have form $\lambda x: code_{f}(x)$  and higher-order method $o.m: (A \Rightarrow B) \Rightarrow C$.. For simplicity, let's assume that $o$ is reference to external symbol and not need cps-transformation itself, since we want to show only function call transaltions here.  Async transformation transform $code:X$ into $C_F[code]: F[X]$, where $F$ is our monad. 

Let us informally describe a set of transformations used to translate function call:
$$
\begin{array}{l l}
  \text{ unchanged\,\,\,\,\,} &
                \frac{C_F \llbracket  o.m(f) \rrbracket ,  C_F(code_f) == F.pure(code_f) } { F.pure(o.m(f)) } \\
  \\
  \text{ monadic }  &
              \frac
              {C_f \llbracket  o.m(f) \rrbracket , B = F[B']  }
              { F.pure(o.m(\lambda x: A \Rightarrow CpsMonad[F].flatMap(C_F \llbracket code_f \rrbracket  ))(identity) } \\            
 \\              
  \text{ asyncShift-fo}  & 
          \frac
          { \begin{array}{l l}
             \scriptstyle{ C_F \llbracket o.m(f) \rrbracket  }, 
                                & \scriptstyle{ \exists asyncShift_O = summon[AsyncShift[O]] : }  \\
                               & \scriptstyle{ asyncShift_O.m:  O\times CpsMonad[F] \Rightarrow (A \Rightarrow F[B]) \Rightarrow F[C]  } \\
            \end{array}        
           } 
           {asyncShift_O.m(monad,o)(x \Rightarrow C_F \llbracket  code_f(x) \rrbracket  }
        \\           
        \\
  \text{ asyncShift-o}  & 
          { \begin{array}{l l}
            C_F \llbracket o.m(f) \rrbracket , & \exists asyncShift_O = summon[AsyncShift[O]] :  \\
                               & asyncShift_O.m:  O \Rightarrow (A \Rightarrow F[B]) \Rightarrow F[C]  \\
            \end{array}        
            }
                    \over 
           {asyncShift_O.m(o)(x \Rightarrow C_F \llbracket  code_f(x) \rrbracket  }
        \\           
        \\
    \text{ inplace-f} & 
        {\frac
        {.C_f \llbracket  o_m(f) \rrbracket , \exists m_{asyncShift}  \in methods(O) : CpsMonad[F] \Rightarrow (A \Rightarrow F[B]) \Rightarrow F[C] }
         { m_{asyncShift}(C_F \llbracket code_f(x) \rrbracket ) }
          }
        \\
        \\
    \text{ inplace} & 
        {\frac
        {.C_f \llbracket  o_m(f) \rrbracket , \exists m_{asyncShift}  \in methods(O) : (A \Rightarrow F[B]) \Rightarrow F[C] }
         { m_{asyncShift}(C_F \llbracket code_f(x) \rrbracket ) }
          }
        \\
        \\
               
 \end{array}
$$

Explanation:
\begin{itemize}
  \item In {\textbf unchanged} case we can leave the call unchanged because no cps transformation was needed. Note, that this handle a case when we have no acccess to the source of the $f$ argument: if $x$ is defined externally it can't contains \lstinline|await|. 
  \item In {\textbf monadic} case is possible to reshape function arguments, to keep the same signature to receive.
  \item case {\textbf asyncShift-fo}  define call substitution:  If we have instance $asyncShift_O$ of marker typeclass  \lstinline|AsyncShift[O]|,  which provide a substitution methods. 
    $m_{shift}$ with additional parameter list where we pass original object and target monad.
    
    |.e. let we have class with higher-order function, for example:
  
\begin{lstlisting}
class  Cache[K,V]  {
   def getOrUpdate(k: K, whenAbsent:  =>V): V
}
\end{lstlisting}  
 and want to use this class in asynchronous environment like next code fragment:
\begin{lstlisting}
async[Future] {
    ...
    cache.getOrUpdate(k, await(fetchValue(k)))
}  
\end{lstlisting} 
   where fetchValue return \lstinline|Future[V]|. 
   
  For defined async substitution for \lstinline|getOrUpdage| method we should define a given instance of marker typeclass $asyncShift_{Cache}$ when shifted method is defined.
\begin{lstlisting}
class CacheAsyncShift[K,V] extends AsyncShift[Cache[K,V]]{

   def getOrUpdate[F[_]](o:Cache[K,V],m:CpsMonad[F])
                    (k:K, whenAbsent: ()=> F[V]):F[V] =
     ....

}
given CacheAsyncShift[K,V]()
\end{lstlisting}   
  Here substitution method have one additional list of arguments \lstinline|(o:Cache[K,V],m:CpsMonad[F])|, where we pass original object itself and our target monad. Since monad parameter is generic, we also have additional type parameter \lstinline|F|.
  Functional call will be transformed to
\begin{lstlisting}
 summon[AsyncCache[Cache[K,V]]]
     .getOrUpdate[Future](cache,monad)(k, ()=>fetchValue(k))
\end{lstlisting}   
   and \lstinline|summon[AsyncCache[Cache[K,V]]]| will be resolved to \lstinline|CacheAsyncShift[K,V]| by implicit resolution rules, so resulting expression will be
\begin{lstlisting}
  CacheAsyncShift[K,V]()
     .getOrUpdate[Future](cache,monad)(k, ()=>fetchValue(k))
\end{lstlisting}   
   \item case {\textbf asyncShift-o} is a modification of the previous rule for the situation when our target monad already parametrizes the substitution class, so we do not need an extra type parameter and monad instance in the additional parameter list.
    \item case {\textbf inplace-f } and {\textbf inplace }  
    describe a situation when the author of a class is aware of the existence of dotty-cps-async and define a shifted
    method in the same scope as the original method.  By convention, such shifted methods are prefixed with \lstinline|Async| suffix.  
    
    Example:  
\begin{lstlisting}
class  Cache[K,V]  {

   def getOrUpdate(k: K, whenAbsent:  =>V): V

   def getOrUpdateAsync[F[_]](m: CpsMonad[F])
               (k: K, whenAbsent: () => F[V]): F[V]

}
\end{lstlisting}

 \end{itemize}
 
   Such substitutors for most higher-order functions from Scala standard library are supplied with dotty-cps-async runtime. Also, developers can provide their substitution for third-party libraries.

  The return type of substituted function can be:
  \begin{itemize}
    \item $C$, the same as the origin
    \item $F[C]$  origin return type wrapped into the monad. 
    \item \lstinline|CallChainAsyncShiftSubst[F,C,F[C]]|. This is a special marker interface for call chain substitution, which wich will be described later.
  \end{itemize}

\begin{itemize}

 \item Exists method in \lstinline|O| with name \lstinline|m_async| or \lstinline|mAsync| which accept shifted argument $f: A \Rightarrow F[B]$. The conventions for the return type are the same as in the previous case.  This case is helpful for the fluent development of API, which is accessible in both synchronous and asynchronous forms. 

 \item If none of the above is satisfied, the macro generates a compile-time error.

\end{itemize}

These rules are extended to multiple parameters and multiple parameters list, assuming that if we have one higher-order async parameter, then all other parameters should also be transformed, having only one representation of the asynchronous method.

\subsection{ Call-chain substitutions }
 As shown in previous section, one of the possible variant of return method of substituted higher-order function is 
  \\
  \lstinline|CallChainAsyncShiftSubst[F[_],B,F[B]]|. 
The developer can use this variant to delay applying $F[\_]$  till the end of the call chain.

For example, let's look at the next block of code:

\begin{lstlisting}
   for { url <- urls if await(score(url)) > limit) }
      yield await(fetchData(url)
\end{lstlisting}
wich is desugared as 
\begin{lstlisting}
  urls.withFilter(
        url => await(score(url)) > limit
       ).map(url => await(fetchData))
\end{lstlisting}

The programmer expects that the behavior of the code should be the same, regardless of using \lstinline|await| inside a loop, so the list of URL-s will be iterated once.  However, if the result of \lstinline|withFilter| has form \lstinline|F[List.WithFilter]|, two iterations are performed - one for filtering the list of URLs and the other over the filtered list to perform fetching data.   User objects for call-chain substitution can accumulate the sequence of higher-order functions in one batch and perform iteration once. 

This block of code is transformed as follows:

\begin{lstlisting}
  summon[AsyncShift[List[String]].withFilter[F](urls,m)(
       url => m.map(score(url))(x=>x>limit)
      )             // CallChainAsyncShiftSubst[F,WithFilter,F[A]]
     .mapAsync(url => fetchData)    // function added to builder
     ._finishChain()                     // finally eval all.
\end{lstlisting}

\subsection{ Automatic coloring }. 
\label{AutomaticColoring}

Automatic coloring is the way to free the developer from writing boilerplate await statements.  Since most industrial code is built with some asynchronous framework, await expressions are often situated literally in each line.  Those expressions do not carry out business logic; when writing code, we should not care how an object is coming to code, synchronously or asynchronously, the same as we do not care how memory to our objects should be allocated and deallocated. 
Apart from performance-critical applications (e.g., developing a web server for which the developer requires complete control of low-level concurrency details), it is preferable to use higher-level constructs hiding low-level details. 
. When writing business logic using some low-level system framework, we expect that framework provides a reasonable generic concurrency model and abstracts away from manual coloring.

 We can provide implicit conversion from $F[T]$ to $T$.  Can we make such conversion safe and preserve semantics with automatic coloring? 
 
 It is safe when $F[\_]$ is a \lstinline|Future|, because multiple calls of \lstinline|await| on \lstinline|Future| produce the same effect as one call -- after the result value will be available after first await, it will be returned immediately after other calls of awaits of the same Future.  I.e., we can say, that \lstinline|Future| is cached.

    For other types of monads, where each \lstinline|await| can perform a new computation, such implicit conversion will be unsafe - the behavor of the following code snippets will be different: 
\begin{lstlisting}
val a = x
f1(await(a))
f2(await(a))
\end{lstlisting}    

and
\begin{lstlisting}
val a = await(x)
f1(a)
f2(a)
\end{lstlisting}    

   To overcome this, we can provide memoization of execution, by embedding the memoization into the transformation of val definitions.  

 Let us have block of code \lstinline[basicstyle=\small]|{ val v = expr; | $tail_v$ \lstinline| }|, $expr$ return value of type $F[T]$ and exists \lstinline|CpsMemoization[F]| with method \lstinline|apply[T](F[T]):F[F[T]]|.

   Cps transformer can check the variable type and rewrite this to.
    $$
      \texttt{summon[CpsMonad[F]]}.\texttt{flatMap}(
              \texttt{CpsMemoization[F]}(expr))( v1 \Rightarrow cps(tail_{v1}) )
    $$
\normalsize

Implicit conversions are often criticized as an unsafe technique, which can be a source of bugs and maintainability problems. In our case, uncontrolled usage of implicit conversions can break the semantics of building complex effects, where some building parts can be automatically memoized.   Dotty-cps-async implements preliminary analysis of automatically generated conversion, which emits errors when detecting potentially unsafe usage.

 To make transformation safe, we should check that developer cannot pass memoized value to API, which expects a delayed effect.  Preliminary analysis ensures that all usages of memoized values are in synchronous context by forcing the next rules:

\begin{itemize}
 \item If some variable is used only in a synchronous context (i.e., via await), the macro will color it as synchronous (i.e., cached if used more than once).
 \item If some variable is passed to other functions as an effect - it is colored as asynchronous (i.e., uncached).
 \item If the variable is simultaneously used in synchronous and asynchronous contexts, we cannot deduce the programmer’s intention, and the coloring macro will report an error.
 \item If the variable, defined outside of the async block, is used in synchronous context more than once - the macro also will report an error.
\end{itemize}

 Behind providing implicit conversion, automatic coloring should also care about value discarding: expressions that provide only side-effects are not an assignment to some value but discarded. When we do automatic coloring, the monad with side-effect generation becomes the value of an expression.  So, we should also transform statements with value discard to insert awaits there.  
Dotty-cps-async interfaces has a \lstinline|ValueDiscard[T]| typeclass.  The statement inside async block can discard value of type \lstinline[basicstyle=\small]|T| only if exists implementation of \lstinline|ValueDiscard[T]| interfaces: in such case macro transforms value discard into \lstinline|summon[ValueDiscard[T]].discard(t)|.

 A special marker typeclass  \lstinline|AwaitValueDiscard[F[T]]| is used when this value discard should be a call to await.  

If we will apply automatic coloring to our example with copying file, we will see that difference between synchronous and asynchronous code become invisible.

\begin{figure}
\label{AACopyFileAC}
\begin{lstlisting}
 val prg = async[[X] =>> Resource[IO,X]] {
   val input = open(Paths.get(inputName),READ)
   val output = open(outputName,WRITE, CREATE, TRUNCATE_EXISTING)
   var nBytes = 0
   while
      val buffer = read(input, BUF_SIZE)
      val cBytes = buffer.position()
      write(output, buffer)
      nBytes += cBytes
      cBytes == BUF_SIZE
   do ()
   nBytes
 }
\end{lstlisting}
\end{figure}

\section{ Related work } \label{RelatedWork}

The idea of 'virtual' program flow encapsulated in a monad is tracked to ({\it{Claessen}} \cite{claessen_1999}), which become a foundation for Haskell concurrent library.  Later \verb|F#| computation expressions were implemented as further development of do-notation. Furthermore, \verb|C#| moves async/await from virtual monadic control-flow to 'normal control-flow,‘  which becomes a pattern for other languages ({\it{Syme}} \cite{10.1145/3386325}). ({\it{Tomas and Syme}} \cite{computation-zoo-padl14}) provides an overview of computation expression usage in different areas.

 Generic monadic operation pairs [reify/reflect] and links between monadic and cps transformations are described in ({\it Filinski } \cite{10.1145/174675.178047}.)

 In scala land, the first cps transformer was implemented as a compiler plugin ({\it Rompf, Maier, Odersky })\cite{DBLP:conf/icfp/RompfMO09}. It provides quite a powerful but complex interface based on delimited continuations. 
  Scala-Async ({\it Haller} \cite{hallerScalaAsync}) provides a more familiar interface for developers for organizing asynchronous processing by compling async control flow to state machines. The main limitation is the absence of exception handling.  Recently, a Lightbend team moved implementation of scala-async from macro to compiler plugin and extended one to support external 'Future systems' such as IO or Monix.
  Although dotty-cps-async is internally based on another type of transformation, it can be viewed as an extension of the scala-async interface for the next language version with a similar role in the Scala ecosystem. The new facilities are a generic monad interface, support of try/catch, and limited support for higher-order functions.
  
In ({\it Haller and Miller } \cite{DBLP:journals/corr/HallerM15})  scala-async model is extended to handle reactive streams. Scala coroutines  ({\it Prokopec and Fengyun} \cite{prokopec_et_al:LIPIcs:2018:9208})  provides a model which allows to build async/await interface on top of coroutines.
  Scala Virtualized ({\it Rompf, Amin, Moors, Haller} \cite{ScalaVirtualized}) devotes to solving a more general problem: providing deep embedding not only for monadic costructions but for arbitrary language.
  Scala Effekt ({ \it Brachth{\"{a}}user and others } \cite{DBLP:journals/jfp/BrachthauserSO20})  allows interpretation of effect handlers inside control monad with delimited continuations.  The same authors released a  monadic reflection library for scala3  ({ \it Brachth{\"{a}}user and others } \cite{brachthaeuser21representing}),   using the capabilities of yet not released support for continuations in JVM in ({\it Project Loom}\cite{ProjectLoom}).  This approach can be a convenient way for implementing async/await like functionality for the future versions of the JVM,  for monads, which can be implemented on top of the one-shot continuations.  Note, that example of combinatorial search from section \ref{MonadsParametrization} on page \pageref{CombSearch},  cannot be implemented with the runtime monadic  because combinatorial search, as other applications of non-determenism requires multiple-shot continuations, where captured continuations can be invoked more than once.

\section{ Conclusion and further work. } \label{Conclusion}

  Prerelease versions of dotty-cps-async have been available as open-source for more than a year, and we have some information based on actual usage in application projects.  Macro library is used in open-source chatbot server with \lstinline|Future| based stack. An experimental proofspace.id internal microservice for connecting PostgreSQL database to VoIP server is built with the help of dotty-cps-async with \lstinline|cats-effect| stack.  The most frequently used monad here is \lstinline|[X] =>> Resource[IO,X]| (the common pattern is to acquire database connection from the pool for each request).  Also, we can point to the port of scala-gopher to scala3, which provides an implementation of concurrent sequential process primitives on top of generic monadic API and experimental TypeLevel project, which brings the direct style into the cats-effect ecosystem.

  Overall feedback is mostly positive.  Reported issues usually have form of inability to compile some specific tree and often
   tracked down to the issues in the compiler. Note, that scala3 compiler also was higthly experimental at this time.  The number of reports about ergonomic and diagnostics is relative low.  Particulary this is because scala3 macros is applied after typing, so usually type errors are catched before macro applications. In some cases we have an error during retyping of transformed tree: for this case the path for better error diagnostics was  was submitted and merged into scala3 compiler.  The work of the patch is extending error message by showing the tree, which was transformed by current macros, in addition to the code  position under -explain compiler option.
   
 Performance issues are not reported at all.  Dotty-cps-async does not provide its own asynchronous runtime but is used with some existing runtime.  Furthermore, if synchronous code remains unchanged and asynchronous code does not add extra runtime operations, then any benchmark will show a performance of the underlying runtime.

  Ability to use direct control-flow on top of some library is a one half of programming experience.  The other part is the library itself.  Currently, we have a set of asynchronous scala runtimes with a different sets of capabilities and it would be interesting to build some uniform facilities for concurrency programming.  One of the open questions is to extend eager Future runtime to support structured concurrency; Problem from the other side -- users of effect stacks, such as IO, need to wrap impure API into effects. Can we automate this process?   

 Also it will be interesting to adopt using of runtime continuations instead compile-time transformations for some type of monads, which will eliminate the need to manually write substitutions for higher-order functions on continuation-enabled platforms. 
     Another direction is the expressivity of internal language, which can be extended by building appropriate wrapper control monad. 

\bibliography{cps}

\end{document}